# Establishing a library of metasurface building blocks through coherence-controlled holographic microscopy

Ondřej Červinka[1,2], Vlastimil Weiss[1], Martin Hrtoň[1,2], Petr Bouchal[1,2], Petr Liška[1,2], Filip Ligmajer[1,2], Tomáš Šikola[1,2], Petr Viewegh[1,2,*]

[1]*Institute of Physical Engineering, Faculty of Mechanical Engineering, Brno University of Technology, Technická 2, 616 69 Brno, Czech Republic*

[2]*Central European Institute of Technology, Brno University of Technology, Purkyňova 656/123, 612 00 Brno, Czech Republic*

*\*petr.viewegh@ceitec.vutbr.cz*

**Abstract**

Digital holographic microscopy is a powerful tool for characterizing transparent and reflective phase objects. Its ability to reconstruct amplitude and phase can also offer great insight into wavefront shaping and design of all-dielectric optical metasurfaces. While metasurfaces have reached widespread popularity, their design is often based purely on the results of numerical simulations which can overlook many of the real-world fabrication imperfections. Being able to verify the real phase response of a fabricated device is of great utility for high-performance devices. Here, we use holographic microscopy to validate metalibraries of rectangular $TiO_2$ and Si building blocks. Illumination effects are studied for wavelengths from 600 to 740 nm and linear polarization rotating within the full range of unique states (0°–180°). Finally, by varying the numerical aperture of the condenser lens from 0.05 to 0.5 we also study the effects of an off-axis illumination. Comparing the experimental results with simulations from finite-difference time-domain and rigorous coupled-wave analysis, we highlight the limitations of these theoretical predictions and underscore the utility of an experimentally established library of building blocks. We demonstrate that our proposed method of holographic microscopy is both practical and effective for creating a metalibrary that accounts for all fabrication and material imperfections, which is crucial for designing high-efficiency metasurfaces.

**Introduction**

Optical metasurfaces represent a promising approach to the fabrication of state-of-the-art free-space optical components and on-chip devices.[1,2] Their large potential stems from the vast design space offered by the subwavelength building blocks.[3,4] In the last decade, the majority of conventional optical components have been reproduced using metal-based plasmonic metasurfaces.[5–7] However, these metasurfaces often provided insufficient efficiency due to the low scattering cross-sections of individual nanoantennas.[8] A significant transformation of the field came with the emergence of dielectric-based optical metasurfaces (e.g. $TiO_2$, Si, $SiO_2$), pushing light-scattering efficiencies over 90 %.[9–13] The progress within the field gave rise to many types of dielectric optical metasurfaces such as Huygens[14,15], Pancharatnam-Berry[12], and others[16,17].

High efficiencies of the final metasurface-based devices rely on closely matching the phase response of individual building blocks to the optimal design, which gets increasingly more complex with additional wavelength-, polarization-, or angle-multiplexing. However, the characterization of optical metasurfaces is often performed only by verification of the intended functionality in the far-field region or by inspection of the intensity distribution in optical microscopes.[3] Such characterization evaluates overall metasurface response but prevents detailed identification of the various sources of functional imperfections. Oftentimes, the sub-optimal performance of a metasurface gets disregarded as "fabrication imperfections" of the building blocks, no matter whether they originate from size deviations, rounding of the edges, incorrect material dielectric function, or even from an incorrectly established library of building blocks. To verify the proper response of the building blocks the fabricated metasurface can be analyzed experimentally.[18] Scattered light amplitude and phase of individual building blocks can be measured directly in the metasurface plane using scattering-type scanning near-field optical microscopy (s-SNOM).[19] Despite the high spatial resolution, such measurements are impractical due to slow imaging speed, small field of view, and overall complexity of the experiment. Hence, digital holographic microscopy has been previously established as a powerful tool for metasurface amplitude and phase visualization, overcoming the limitations of near-field techniques.[20,21] Holographic microscopy offers non-destructive imaging over a large field of view, while maintaining high lateral resolution. Demonstrated experiments are directly applicable to various types of metasurfaces, including those with metallic[22], dielectric[21], or geometric-phase building blocks[20]. From the portfolio of available holographic techniques, we found coherence-controlled holographic microscopy (CCHM) very advantageous,[22,23] as it provides amplitude and phase imaging in a single snapshot using the light of arbitrary spatial and temporal coherence. Single-shot imaging might be necessary for studying active metasurface in the future,[24–26] but it also minimizes the influence of external disturbing stimuli on amplitude and phase reconstruction. The inherent control over spatial coherence allows us to study the building blocks not only under normal incidence of plane waves but also under oblique illumination.[27] Further, the inherent achromaticity of the interferometer and its robustness to temporally incoherent light allow measurements at a broad range of illumination wavelengths. All these properties enable us to study metasurface components and their building blocks in full complexity and under practical conditions.

Here, we demonstrate CCHM as a robust and powerful tool for establishing a metasurface building blocks library, focusing on anisotropic high-aspect-ratio $TiO_2$ building blocks as a case study. We showcase how CCHM enables characterization of the collective response of the building blocks with respect to the changes in illumination parameters such as wavelength (600 to 740 nm), linear polarization (0° to 180°), and numerical aperture (0.05 to 0.5). Comparing experiments and numerical simulations, we identify discrepancies between true optical response and theory, shedding light on the challenges and opportunities in metasurface design and research.

**Results and Discussion**

The experimental setup of CCHM is depicted in Fig. 1a. The system is derived from a commercially available Q-Phase microscope (Telight company), which is based on an achromatic off-axis Mach-Zehnder interferometer. For the purpose of the manuscript, we replaced the original light source with a monochromator and inserted a linear polarizer into the illumination path. The light from the monochromator was coupled to the fiber bundle (FB), representing the area source at the microscope input. The condenser lens aperture was then used to control the apparent size of the source and thus the illumination numerical aperture. The light from the fiber bundle passed through a rotatable linear polarizer (P) and illumination lens (IL), and was divided into the signal and reference arms by the beamsplitter (BS). It was then reflected at the plane mirrors (M), transformed by condenser lenses (C), and it finally illuminated the sample (S) and the reference sample (RS). We worked with the same microscope objectives (MO, 10×, NA 0.3) in both arms. The light was coherently recombined at the CCD using tube lenses (TL), beam splitters (BS), diffraction grating (DG), and camera lenses (CL). Individual wavelengths interfered at slightly different angles, creating the same carrier frequency in holograms recorded at different wavelengths. Hence, the hologram amplitude and phase reconstructions at individual wavelengths were well-comparable to each other. More information on hologram reconstruction is in Methods.

To experimentally establish the building blocks library, we have prepared a sample composed of 20×20 μm fields, each filled with an array of anisotropic TiO$_2$ building blocks of different widths *W* and lengths *L* ranging from 130 to 330 nm (Fig. 1b,c). The range of nanostructure sizes was chosen because it is manufacturable with our lithography system and also the RCWA simulations predict it to cover the full 2π range. The pitch *P* = 450 nm and height *H* = (500 ± 10) nm of all blocks were kept constant. The blocks were fabricated on a fused silica substrate using electron beam lithography (see Methods). In Fig. 1d, we show a white-light optical image of the fabricated sample. The fabricated height and periodicity were verified by atomic force microscope (AFM, see the inset of Fig. 1e). With the precise dimensions of the nanostructures analyzed, we can now move on to the subsequent experiments with CCHM, focusing on the amplitude and phase response of the nanostructure arrays. In the following sections, we discuss the effects of wavelength, polarization and numerical aperture, compare the experimental results to numerical simulations based on the measured fabricated dimensions of the nanostructures.

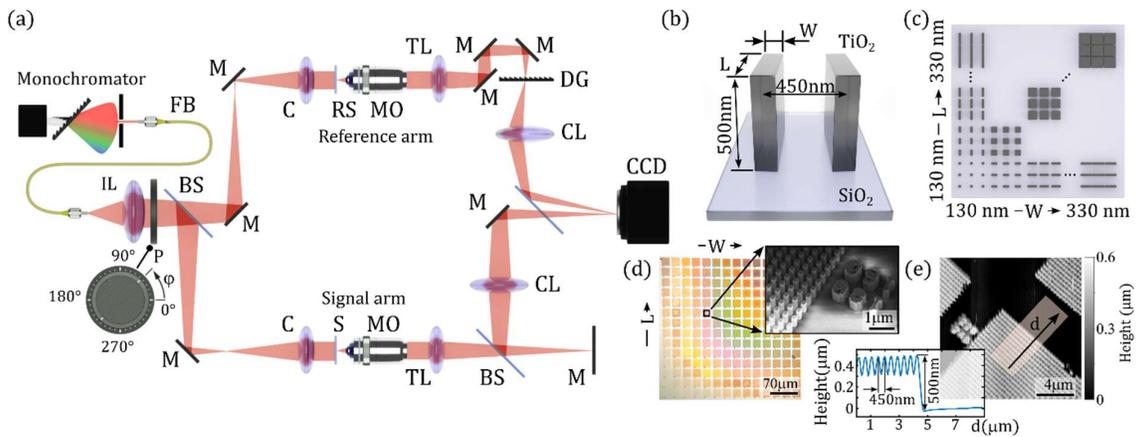

*Fig 1: (a) Experimental setup used in holographic measurements: monochromator varying illumination wavelength, FB – fiber bundle, IL – illumination lens, P – rotating linear polarizer, BS – beam splitter, M – plano mirror, C – condenser, RS – reference sample, S – sample, MO – microscope objective, TL – tube lens, DG – diffraction grating, CL – camera lens. (b) Scheme of TiO$_2$ blocks with width W, and length L fabricated on SiO$_2$ substrate. (c) Illustration how the width W and length L of building blocks vary between the fields on the sample. (d) White-light microscope image of the sample from (c) with SEM detail of actual building blocks. (e) AFM image of the sample with topographic profile verifying the height of fabricated blocks. The oscillations in the AFM measurement are caused by the AFM tip sliding over the top and partially in between the nanostructures, so only the top value corresponds to the true height of the nanostructures.*

Our first CCHM experiment demonstrates the dependence of the scattered light phase and amplitude on the illumination wavelength. We recorded holograms of the sample for the wavelength sweep by tuning the monochromator from 600 to 740 nm with 5 nm step. The light was linearly polarized in the horizontal direction (compare arrows in Fig. 2 with Fig. 1c). The *phase* reconstructed for all sample fields at three representative illumination wavelengths (600, 670, and 740 nm) is shown in Fig. 2a, while the rest of the data at other wavelengths is available in Supplementary materials (Fig. S1). Although each field exhibited minimal variance in the recorded phase values, the plotted values are for clarity averaged over the full field of nanostructures, excluding the area of the first few rows of nanostructures from the edges of each field. The full spectral dependence of the phase for four selected fields is shown in Fig. 2b. The reconstructed *amplitude* at the same three wavelengths is shown in Fig. 2e (with the rest of the data available in Supplementary materials, Fig. S2). The full spectral dependence of the amplitude is shown in Fig. 2f, analogously to the phase in Fig. 2b. Following these experiments, we also conducted rigorous coupled-wave analysis (RCWA) and finite-difference time-domain (FDTD) (Supplementary materials, Figs. S3) simulations of the scattered amplitude and phase for the building blocks closely representing the fabricated ones. The phase (Fig. 2c,d) and amplitude (Fig. 2g,h) trends from numerical simulations align well with the experimental results despite the discrepancies in absolute values. The experimental phase values saturated at $1.25\pi$, while the values in numerical simulations reached up to $2\pi$. Even though the reduced phase coverage severely limits the forward design of metasurfaces, the recently developed inverse design algorithms can produce an efficient design even without the full 0-$2\pi$ phase coverage [28,29]. Alternatively, this finding could be used to extend the parameter sweep of the building block parameters used. Similar to Refs. [30,31], we achieved better agreement between experiment and simulations for lower phase values, while the differences were greater for larger phase shift values. We investigated whether this discrepancy could have been caused by modification of the $TiO_2$ dielectric function during the fabrication process. While the simulations were performed using the dielectric function obtained through ellipsometry on 150 nm ALD $TiO_2$ layer shortly after the deposition, during the fabrication the nanostructures went through an additional process of $O_2$ plasma (during the resist ashing). To answer the question of possible material modification, we also measured the dielectric function of the $TiO_2$ layer after the $O_2$ plasma exposure and observed that the real part of the dielectric function actually increased slightly. In high-aspect-ratio nanostructures, which have a naturally larger portion of surface exposed to plasma, this effect would be even more pronounced, which would lead to an increase in measured phase delay and cannot, therefore, explain the discrepancy. Other effects tested, like the rounding of the edges and imprecisely measured nanostructure height, also showed no meaningful modification of the resulting data. We finally concluded that the most likely explanation is a formation of hollow spaces within the nanostructures, caused by the limited diffusion of ALD precursors into the high-aspect-ratio holes in the resist and slightly positive sidewall angle of some nanostructures. Such fabrication defects are not directly observable by non-destructive analysis techniques and would go unnoticed without the phase shift verification by the CCHM, showing another advantage of composing the building block library experimentally. To verify our hypothesis, we used a focused ion beam to cut several different-sized nanostructures in half to directly see their cross-section. While the smallest nanostructures were, to the limit of the measurement, filled completely, the larger nanostructures displayed a significant void in their center, as can be seen in Fig S4. Simulating the voids inside nanostructures reveals a substantial decrease in phase shift provided by the nanostructure and also shifts the spectral position of the amplitude dips. Because the void regions varied greatly among the nanostructures and were thus difficult to parametrize, we did not include them in any of the further shown simulations.

Experiments and numerical simulations both attest that the phase is more sensitive to changes in block dimensions parallel with the illumination polarization regardless of the illumination wavelength. Plots in Fig. 2b further demonstrate negligible phase dispersion of small nanostructures, while the larger nanostructures exhibit a linear decrease in phase with a slope of around –5.6 mrad/nm. These values give an estimate of phase errors when a metasurface is illuminated with wavelengths other than the designed one. The simulated amplitudes also qualitatively agree with the simulations, especially in the areas of decreased amplitude attributed to resonances within the nanostructures. The transmittance of the nanostructures changes more rapidly with wavelength when the short axis of the nanostructures is parallel with the illumination polarization (see Fig. S2), showing that the modes excited in the nanostructure are much more dependent on the short axis dimension. Overall, the numerical simulations exhibit much sharper features in phase and dips in transmittance when compared to experimental results. However, these features are smoothed out when minimal extra absorption (k=0.002) was introduced into the simulated $TiO_2$ material, leading to a better agreement with the experiment (see FDTD simulations in Fig. S3). Note that a similar effect would be achieved by breaking the perfect periodicity of the arrays, similar to non-zero surface roughness of nanostructures in real-world samples, limiting the effect of rod-rod coupling originating from periodic boundary conditions, or by limiting the spatial extent of the nanostructure arrays in the simulations.

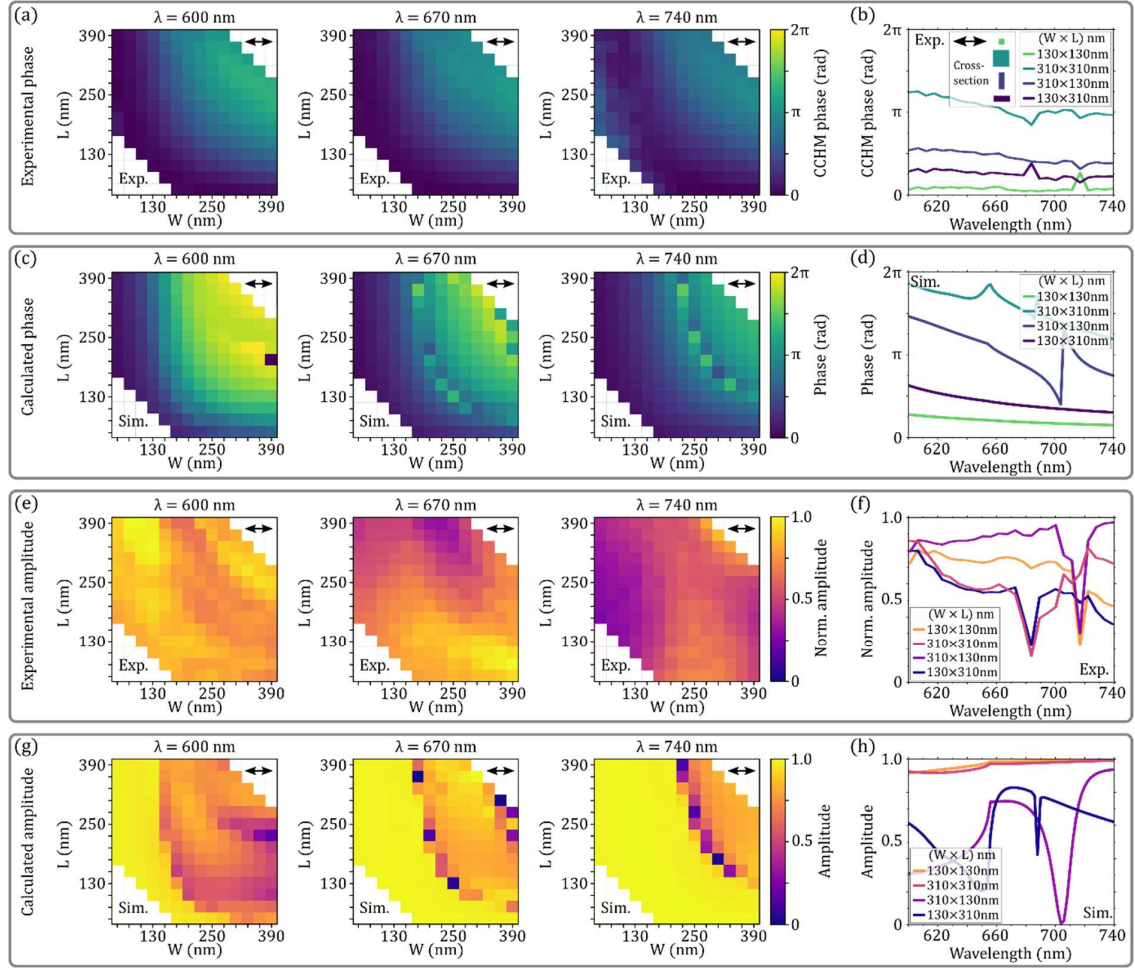

*Fig 2: (a) CCHM phase reconstructed for all fields at three selected wavelengths and (b) for selected fields at full spectrum. (c) RCWA simulated phases for all fields at three selected wavelengths and (d) for selected fields at full spectrum. Similarly, CCHM reconstructed norm. amplitude (e, f) and RCWA simulated amplitude response (g, h) for various illumination wavelengths. The black arrow in each figure indicates the orientation of the linear polarization of the illumination relative to the orientation of the building blocks.*

The second CCHM experiment investigated how the amplitude and phase of scattered light depend on the polarization of the incident illumination (at the 660 nm wavelength). The illumination was linearly polarized prior to division into signal and reference arms of the CCHM, and polarization was gradually rotated within the full extent of all unique polarization states (0°-180°). The phase reconstructions of the entire sample for horizontal, diagonal, and vertical polarizations are presented in Fig. 3a. The phases of remaining polarizations are available in Supporting information, Fig. S5. Phase reconstruction with fine increments of the illumination polarization angle is demonstrated in Fig. 3b for selected nanostructure sizes. The non-trivial phase response of the nanostructures as a function of a polarization angle is well described within our theoretical model (cf. Eqs. (4) and (5) in Methods). The amplitude reconstructions of the entire sample for the same three polarizations are presented in

Fig. 3e (see the rest in Supporting information, Fig. S6). Amplitude reconstruction depending on the illumination polarization is demonstrated in Fig. 3f. In line with theoretical expectations, the phase of all square blocks is insensitive to polarization variations (the main diagonal in Fig. 3a). This is also documented by constant phase levels corresponding to square blocks in Fig. 3b. When the polarization is changed from horizontal to vertical, the phase maps are mirrored about the diagonal fields with square blocks. The amplitude, on the other hand, fails to follow this rule, leading to a clear asymmetry at 45° that stands in stark contrast to the expectations based on the symmetry of the problem. Some of the asymmetry can be attributed to polarization-sensitive optical elements, as our instrument was not originally designed to work with polarized light. For example, the diffraction grating was oriented to maximize transmission in the horizontal polarization state, which led to about a 10 % decrease in transmission for the vertical polarization state. Comparing the phase difference between horizontal and vertical polarizations, we can evaluate the retardance of the sample. This is a very important quantity for the design of Pancharatnam-Berry metasurfaces, as the conversion efficiency depends heavily on the retardance provided by the anisotropy of the nanostructure.[32] We can again see qualitative agreement between the experiment and simulation (Fig. 3d). However, the nanostructures predicted to be half-wave plates (1π phase difference) turn out to be only quarter-wave plates (0.5π phase difference) in the experiment, due to the lower phase shift accumulated for each polarization. We also attribute the origin of the decreased retardance to the hollow region within the nanostructures. The simulations for orthogonal linear polarization states at a wavelength of 660 nm in Fig. 3c,g (phase and amplitude) show the same quantitative disagreement as previously discussed.

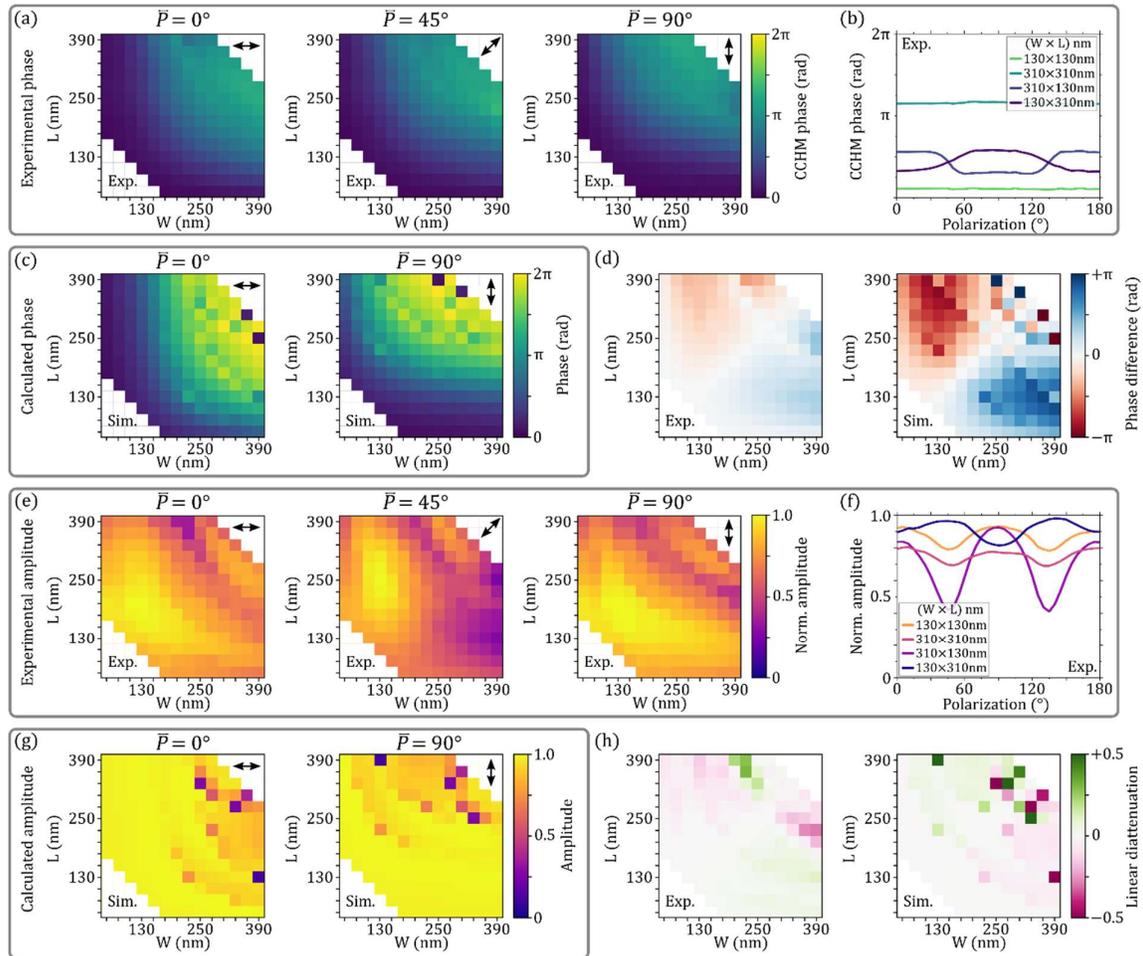

*Fig 3: (a) CCHM phase reconstructed for all fields at three distinct linear polarization angles and (b) for selected*

*fields at finely varied polarization angles. (c) RCWA simulated phases for two orthogonal linear polarization states. (d) Retardance calculated from the measured data and simulations. Similarly, CCHM reconstructed norm. amplitude (e, f) and RCWA simulated amplitude response (g, h) for varying polarizations. The black arrow in each figure indicates the orientation of the linear polarization of the illumination relative to the orientation of the building blocks. All data acquired for wavelength of 660 nm.*

The third CCHM experiment aims to unravel the dependence of the phase (Fig. 4a) and amplitude (Fig. 4c) on the inclined illumination. In numerical simulations, it is customary to assume that incident light is a plane wave with a propagation direction perpendicular to the metasurface. However, the conditions in practical situations can be considerably different. The inclined waves are incident on metasurface lenses during imaging[33] or in applications of wide-angle metasurfaces with various functionalities.[34] Operating CCHM with the condenser iris nearly closed (condenser numerical aperture NA = 0.05), the incident light resembles plane wave illumination. However, the CCHM condenser iris opening can be changed, providing numerical apertures up to 0.5. In this situation, the sample is illuminated by many plane waves simultaneously and the amplitude and phase images of the sample are created as a superposition of measurements at different illumination angles. To study the connection between inclined illumination and the overall sample response, we gradually increased the numerical aperture of the condenser lens from 0.05 to 0.5 in 0.05 steps. The phase response of the nanostructures across the majority of the observed fields remained the same up to NA = 0.5 (Fig. 4a,b and Fig. S7). The amplitude, however, depends dramatically on the numerical aperture (Fig. 4c,d), indicating some of the new angles available with increasing numerical aperture interact very differently with the nanostructures, perhaps because of the voids within them, which leads to a decrease in the intensity of transmitted light. This effect is more prominent in nanostructures whose long axis is aligned with the polarization of light, as can be seen in Fig. S8. The nanostructure dimension perpendicular to the light polarization plays a negligible role in the final response of the nanostructure. Overall, we show that the rarely discussed effect of numerical aperture plays a significant role in the amplitude of transmitted light only.

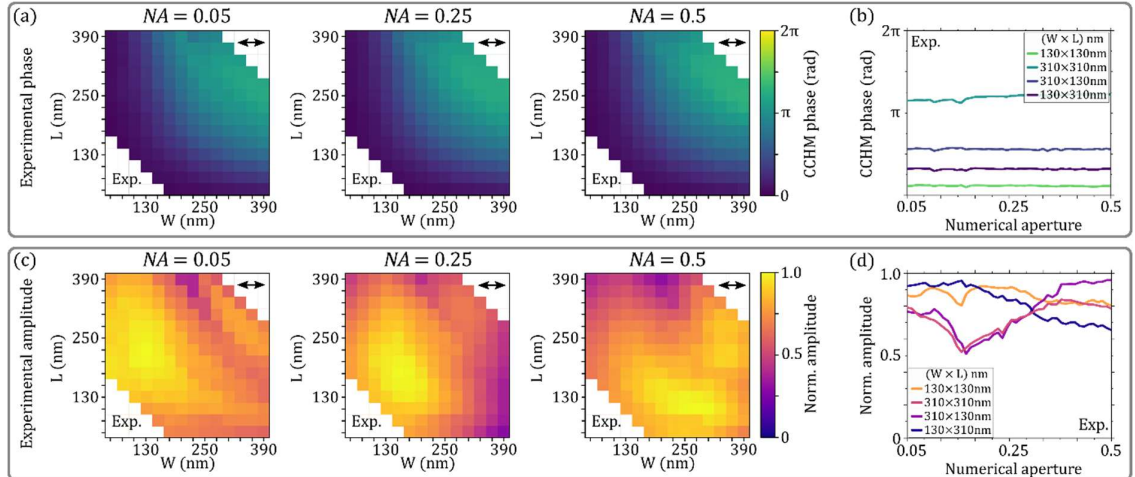

*Fig 4: (a) CCHM phase and (c) norm. amplitude distribution for three numerical apertures. (b) CCHM phase and (d) norm. amplitude dependence for finely varying numerical aperture for select nanostructure sizes.*

Finally, to capture the fine features of resonances and compare them with RCWA and FDTD simulations, we investigate the phase and amplitude properties of polarization-insensitive square nanostructures, increasing in size in 2 nm steps. Fig. 5a shows the phase response as a function of the side length of the square nanostructures with a similar trend between the experiment and the simulations, but an overall lower phase gradient in the experiment. The sharp resonant features observed in the simulations are not present in experiments, indicating that they are not highly relevant to real-world fabricated samples. Fig. 5b shows the transmittance of the same square nanostructures, compared to confocal spectroscopy measurements. Even though both experiments (CCHM and confocal spectroscopy (see Fig. S9)) agree quantitatively, and the two simulation methods (RCWA and FDTD) have a good agreement (except for a small constant offset in phase, which is not relevant in metasurface design), there is still a disparity between the experiments and simulations most likely caused by the presence of voids inside the nanostructures. The resonant features of nanostructures are heavily suppressed even in the transmittance plots owing to fabrication imperfections and finite array area. The lack of resonant features in experimentally acquired phase and transmittance plots is good news for the design of metasurfaces as the rapidly changing values of the resonant features seen in simulations no longer need to be masked or interpolated within the experimentally acquired building block libraries.

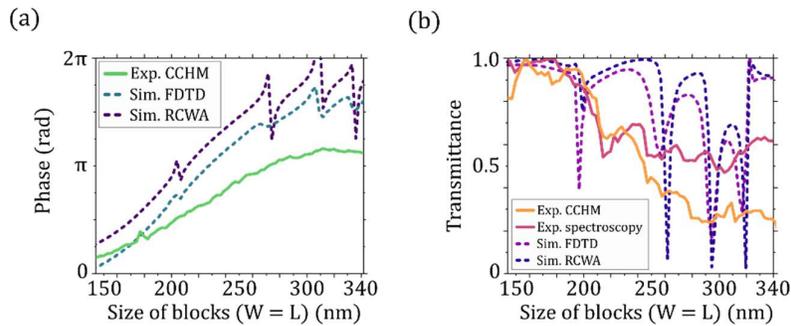

*Fig 5: (a) CCHM phase and (b) norm. transmittance distribution for nanostructures with finely varying square cross-section compared with simulations.*

**Conclusion**

In summary, we demonstrated how experimentally establishing a library of metasurface building blocks using holographic microscopy can reveal significant discrepancies between predicted and actual responses of real-world fabricated samples. We illustrated this workflow by preparing an experimental library of TiO2 building blocks with varied widths and lengths at a constant height. The response of building blocks was analyzed with respect to illumination wavelength, incident light polarization, and numerical aperture. We compared the experimentally obtained results with traditional numerical simulations and discussed the possible sources of deviations previously identified also by other authors. This experimental library underscores the importance of accounting for fabrication imperfections that are overlooked or difficult to quantify in simulations, and it also simplifies metasurface design by eliminating the need to optimize the fabrication process to achieve ideal nanostructure dimensions and shapes. Compared to previously used techniques like QLSI, our proposed method of CCHM does not suffer any major numerical artifacts of phase reconstruction and is suitable for a wide range of metasurface architectures.

Our work provides a robust framework for future metasurface design and fabrication, paving the way for more accurate and reliable metasurface devices. In our vision, simulations will still be used to approximate the range of building block dimensions necessary to cover the full 0-2π phase space. However, each research group experimentally establishes its own building block library via DHM, which encompasses all their fabrication-specific imperfections and design process parameters. The true-response experimentally acquired building block library is then used for the design of the final metasurface devices. Future work might focus on extending the operational range of the instrument into the whole visible spectrum.

## Methods

In optical holography, amplitude and phase reconstruction of the object wave $\vec{E}_o(x,y)$ is facilitated by its interference with a reference wave $\vec{E}_r(x,y)$ in the image plane of the microscope. Denoting $f_c$ the spatial carrier frequency of interference fringes, intensity distribution within the image plane can be written down as

$$I(x,y) = \left|\vec{E}_o(x,y)\right|^2 + \left|\vec{E}_r(x,y)\right|^2 + \vec{E}_o^{\,*}(x,y) \cdot \vec{E}_r(x,y)e^{i2\pi f_c x} + \vec{E}_o(x,y) \cdot \vec{E}_r^{\,*}(x,y)e^{-i2\pi f_c x}. \quad (1)$$

Note that for the sake of simplicity, we set the orientation of the interference fringes to be parallel to the $y$ axis, but it is, in fact, determined by the actual orientation of the diffraction grating inserted into the reference arm of the holographic microscope. Let us further assume that the source wave is linearly polarized and the optical properties of the sample can be represented by a spatially dependent Jones matrix

$$\overleftrightarrow{T}(x,y) = \begin{bmatrix} t_x(x,y) & 0 \\ 0 & t_y(x,y) \end{bmatrix}, \quad (2)$$

where $t_x(x,y)$ and $t_y(x,y)$ are the complex transmission amplitudes along the two principal axes of our metasurface building blocks. Then the cross-term $OR^*(x,y) = \vec{E}_o(x,y) \cdot \vec{E}_r^{\,*}(x,y)$ utilized in the object wave reconstruction (the rest is filtered out during post-processing) becomes

$$I_{or^*}(x,y) = \left[t_x(x,y)\cos^2\theta + t_y(x,y)\sin^2\theta\right] E_r^{\,*}(x,y), \quad (3)$$

where $\theta$ denotes the angle between the incident linear polarization and the $x$ axis of our coordinate system. Careful implementation of background subtraction allows us to completely suppress the effects of the reference wave $E_r^{\,*}(x,y)$, hence the reconstructed amplitude $A(x,y)$ and phase $\varphi(x,y)$ read

$$A(x,y) = |t_x(x,y)|^2 \cos^4\theta + 2\mathrm{Re}\{t_x^{\,*}(x,y)\,t_y(x,y)\}\sin^2\theta\cos^2\theta + |t_y(x,y)|^2 \sin^4\theta, \quad (4)$$

$$\varphi(x,y) = \tan^{-1}\frac{\mathrm{Im}\{t_x(x,y)\,\cos^2\theta + t_y(x,y)\,\sin^2\theta\}}{\mathrm{Re}\{t_x(x,y)\,\cos^2\theta + t_y(x,y)\,\sin^2\theta\}}. \quad (5)$$

The above equations clearly show that the measured amplitude and phase generally correspond to a mixture of the two complex transmission amplitudes $t_x(x,y)$ and $t_y(x,y)$, unless the incident polarization is aligned with one of the building block's principal axes. In the case of unpolarized illumination, the angular dependence completely disappears, and the two complex transmission amplitudes can no longer be separated.

It should be noted that the particularly simple form of the transmission matrix in Eq. (2) stems from the fact that all our building blocks are oriented in the same direction. Also, any anisotropy present in the microscope (e.g. due to the diffraction grating in the reference arm) can introduce additional mixing of $t_x(x,y)$ and $t_y(x,y)$ that is not captured in the formulas above.

As for the experimental realization of our holographic measurements, the input beam of the targeted wavelength from a supercontinuous white source is first selected by a monochromator and then is divided by a beam splitter into two arms (object and reference ones). The light in the object arm interacts with an observed sample, the light in the reference arm passes just through a reference sample. Both beams interfere on the camera plane at the off-axis angle $\vartheta$. When the hologram is obtained using an achromatic and space-invariant off-axis setup, the image term $OR^*$ and twin image term $O^*R$ are separated from the zero-order (autocorrelation) term $|O|^2+|R|^2$ by a spatial carrier frequency $f_c$. The first step of hologram reconstruction is a calculation of the hologram Fourier transform which turns it into the spatial frequency domain, as it is shown in Fig 6. The next step is the extraction of the image term ($OR^*$) using the *circ* function followed by apodization using the Hann window. After shifting the object term to the center of the frequency domain space the complex amplitude $O$ of the object wave can be calculated using the inverse Fourier transform. The phase can be then simply calculated as

$$\varphi = \tan^{-1} \frac{\text{Im}\{O\}}{\text{Re}\{O\}}. \tag{6}$$

In contrast to another commonly used off-axis microscopes, the CCHM uses an incoherent light source (i.e. a halogen lamp with a monochromator), which requires the light path in the object and reference arm to be almost identical (with errors at most in the order of coherence length) to preserve interference, when beams are combined. On the other hand, CCHM can measure dependence of phase and amplitude distribution for various wavelengths of illumination and moreover, incoherent illumination provides better phase resolution due to the lack of any coherence noise.

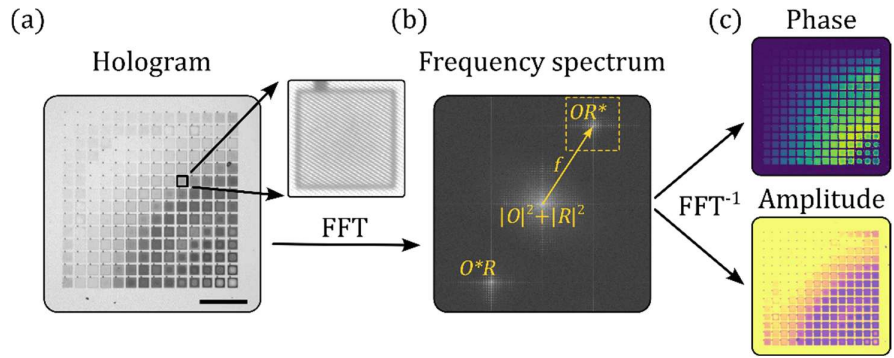

*Fig. 6: (a) Raw hologram of metasurfaces by CCHM. (b) Reconstructed frequency spectrum using Fast Fourier transform (FFT), where we observe three distinct patterns corresponding to zero ($|O|^2+|R|^2$) and +/- first (O*R and OR*) order. (c) Resulting phase and amplitude distributions extracted by inverse FFT.*

**Simulations**

Both RCWA and FDTD simulations were performed using a commercial version of Ansys Lumerical FDTD. The dimensions of the simulated nanostructures were selected to precisely match the dimensions of the fabricated nanostructures measured in SEM with a height of 500 nm measured via AFM. The real and imaginary parts of the refractive index of $TiO_2$ used in simulations were measured through ellipsometry on a 150 nm ALD layer and can be seen in Fig. S10. Finally, the acquired phase shift values of the nanostructures were reduced by the free space propagation phase to provide the same reference point as the experiment.

**Fabrication**

The high aspect ratio TiO$_2$ nanostructures were fabricated using a method combining electron beam lithography with atomic layer deposition similar to Ref. [12]. First, a fused silica wafer (10×10 mm chip) is spin-coated with a resist (Fig. 7(a)). We used a CSAR resist for its good resolution and strong adhesion capabilities. Because the thickness of the resist determines the final height of the nanostructures, it is necessary to select a viscous enough resist. The used CSAR 6200.13 results in a 650 nm layer if spin-coated at 1600 RPM. An additional thin layer of chromium (20 nm) was deposited on top to serve as a charge dissipation layer. The nanostructure layout is then transferred into the resist via electron beam lithography (Fig. 7(b)). We utilized a 30 kV e-beam system RAITH 150 Two for the exposure, with a dose of 200 µC/cm$^2$. The exposed resist was then stripped of the chromium layer and developed using a standard developer amyl acetate for the duration of one minute. The developed resist mask was then transferred into an atomic layer deposition chamber, where it was conformally filled with a 170 nm TiO$_2$ layer (Figs. 7(c),(d)). The ALD recipe was optimized for 90 °C not to melt the e-beam resist mask. 2600 cycles of alternating tetrakis(dimethylamino)titanium and water precursors deposit the 170 nm layer in about 13 hours. The ALD deposition, however, not only fills the resist mask hollows but also creates a layer over the whole sample surface that needs to be removed. To remove the top layer, we etched the sample surface by an argon ion beam in Scia Systems Coat 200 (Fig. 7(e)). This system is also equipped with a secondary ion mass spectrometer, which allows us to stop the etching precisely when the top layer is just removed. In the final step, the sample is put into a Diener resist stripper, where the resist mask is removed in oxygen plasma within 50 minutes (Fig. 7(f)). This leaves the high aspect ratio TiO$_2$ nanostructures free-standing on the substrate. Scanning electron microscope (SEM) images revealed that the fabricated nanostructure dimensions deviate from the designed ones proportionally to the equation $(W,L) = 1.25 \cdot (W_0,L_0)+6$ [nm], meaning the intended dimensions $(W_0,L_0)$ = (40–300) nm were fabricated as $(W,L)$ = (60–380) nm (for more details see Supplementary materials, Fig. S11). The SEM image in Fig. S11 also shows that the blocks have a very low sidewall angle. Using SEM imaging we further found out that the smallest and largest nanostructures were not successfully fabricated. For this reason, they are excluded from further analysis, and the analyzed fields are limited to those with dimensions $(W,L)$ = (130–330) nm. In addition, we also performed measurements of all fabricated nanostructures using confocal optical spectroscopy and the obtained spectra are shown in Fig. S12.

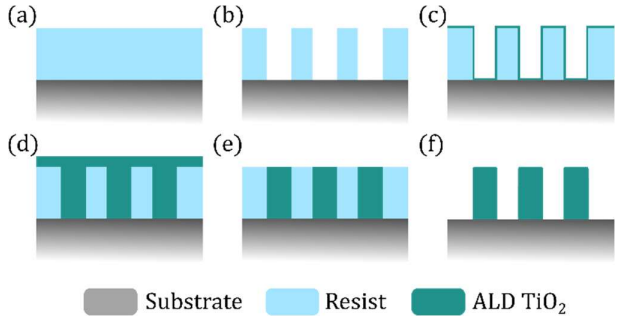

*Fig. 7: (a) Spin-coating the resist. (b) Electron beam lithography and development. (c) Initial stage of ALD TiO$_2$ deposition. (d) Finished ALD deposition. (e) Removal of the top layer by ion beam etching. (f) Removal of the resist by plasma ashing.*

**Acknowledgements**

This work was supported by the Grant Agency of the Czech Republic (20-01673S), and Brno University of Technology (FSI-S-23-8336 and CEITEC VUT-J-24-8583). This work was supported by the project

Quantum materials for applications in sustainable technologies (QM4ST), funded as project No. CZ.02.01.01/00/22_008/0004572 by OP JAK, call Excellent Research. We also acknowledge CzechNanoLab Research Infrastructure supported by MEYS CR (LM2018110) and the Biophotonics Core Facility, CEITEC Brno University of Technology, Czech Republic supported by MEYS CR (LM2023050 Czech-BioImaging).

**Supporting Information**

The Supporting Information is available free of charge at http://pubs.aip.org

Phase response over a wider spectral range (Fig. S1); amplitude response over a wider spectral range (Fig. S2); FDTD simulations of phase and amplitude response (Fig. S3); SEM images of nanostructure cross-sections (Fig. S4); phase response to finer linear polarization angles (Fig. S5); amplitude response to finer linear polarization angles (Fig. S6); phase response to finer numerical aperture (Fig. S7); amplitude response to finer numerical aperture (Fig. S8); confocal microscopy transmittance measurement for finely varying square nanostructures (Fig. S9); ellipsometry measurement of complex refractive index of $TiO_2$ (Fig. S10); a correlation between designed and measured nanostructure size (Fig. S11); transmittance measurement over the whole visible spectrum (Fig. S12)

**Data availability statement**

The data that supports the findings of this study are available from the corresponding author upon reasonable request.

**References**

(1)   Wang, Z.; Li, T.; Soman, A.; Mao, D.; Kananen, T.; Gu, T. On-Chip Wavefront Shaping with Dielectric Metasurface. *Nature Communications* **2019**, *10* (1), 3547. https://doi.org/10.1038/s41467-019-11578-y.

(2)   Li, N.; Xu, Z.; Dong, Y.; Hu, T.; Zhong, Q.; Fu, Y. H.; Zhu, S.; Singh, N. Large-Area Metasurface on CMOS-Compatible Fabrication Platform: Driving Flat Optics from Lab to Fab. *Nanophotonics* **2020**, *9* (10), 3071–3087. https://doi.org/10.1515/nanoph-2020-0063.

(3)   Chen, W. T.; Zhu, A. Y.; Capasso, F. Flat Optics with Dispersion-Engineered Metasurfaces. *Nat Rev Mater* **2020**, *5* (8), 604–620. https://doi.org/10.1038/s41578-020-0203-3.

(4)   Yu, N.; Capasso, F. Flat Optics with Designer Metasurfaces. *Nat Mater* **2014**, *13* (2), 139–150. https://doi.org/10.1038/nmat3839.

(5)   Meinzer, N.; Barnes, W. L.; Hooper, I. R. Plasmonic Meta-Atoms and Metasurfaces. *Nature Publishing Group* **2014**, *8* (12), 889–898. https://doi.org/10.1038/nphoton.2014.247.

(6)   Kildishev, A. V.; Boltasseva, A.; Shalaev, V. M. Planar Photonics with Metasurfaces. *Science* **2013**, *339* (6125), 1232009–1232009. https://doi.org/10.1126/science.1232009.

(7) Giannini, V.; Fernández-Domínguez, A. I.; Heck, S. C.; Maier, S. A. Plasmonic Nanoantennas: Fundamentals and Their Use in Controlling the Radiative Properties of Nanoemitters. *Chemical Reviews* **2011**, *111* (6), 3888–3912. https://doi.org/10.1021/cr1002672.

(8) Yu, N.; Genevet, P.; Kats, M. A.; Aieta, F.; Tetienne, J.-P.; Capasso, F.; Gaburro, Z. Light Propagation with Phase Discontinuities: Generalized Laws of Reflection and Refraction. *Science* **2011**, *334* (6054), 333–337. https://doi.org/10.1126/science.1210713.

(9) Khorasaninejad, M.; Chen, W. T.; Devlin, R. C.; Oh, J.; Zhu, A. Y.; Capasso, F. Metalenses at Visible Wavelengths: Diffraction-Limited Focusing and Subwavelength Resolution Imaging. *Science* **2016**, *352* (6290), 1190–1194. https://doi.org/10.1126/science.aaf6644.

(10) Wang, Y.; Chen, Q.; Yang, W.; Ji, Z.; Jin, L.; Ma, X.; Song, Q.; Boltasseva, A.; Han, J.; Shalaev, V. M.; Xiao, S. High-Efficiency Broadband Achromatic Metalens for near-IR Biological Imaging Window. *Nature Communications* **2021**, *12* (1), 5560. https://doi.org/10.1038/s41467-021-25797-9.

(11) Wu, Y.; Yang, W.; Fan, Y.; Song, Q.; Xiao, S. $TiO_2$ Metasurfaces: From Visible Planar Photonics to Photochemistry. *Science Advances* **2019**, *5* (11). https://doi.org/10.1126/sciadv.aax0939.

(12) Devlin, R. C.; Khorasaninejad, M.; Chen, W. T.; Oh, J.; Capasso, F. Broadband High-Efficiency Dielectric Metasurfaces for the Visible Spectrum. *Proceedings of the National Academy of Sciences* **2016**, *113* (38), 10473–10478. https://doi.org/10.1073/pnas.1611740113.

(13) Arbabi, A.; Horie, Y.; Bagheri, M.; Faraon, A. Dielectric Metasurfaces for Complete Control of Phase and Polarization with Subwavelength Spatial Resolution and High Transmission. *Nature Nanotechnology* **2015**, *10* (11), 937–943. https://doi.org/10.1038/nnano.2015.186.

(14) Leitis, A.; Heßler, A.; Wahl, S.; Wuttig, M.; Taubner, T.; Tittl, A.; Altug, H. All-Dielectric Programmable Huygens' Metasurfaces. *Advanced Functional Materials* **2020**, *30* (19). https://doi.org/10.1002/adfm.201910259.

(15) Decker, M.; Staude, I.; Falkner, M.; Dominguez, J.; Neshev, D. N.; Brener, I.; Pertsch, T.; Kivshar, Y. S. High-Efficiency Dielectric Huygens' Surfaces. *Advanced Optical Materials* **2015**, n/a--n/a. https://doi.org/10.1002/adom.201400584.

(16) Kuznetsov, A. I.; Brongersma, M. L.; Yao, J.; Chen, M. K.; Levy, U.; Tsai, D. P.; Zheludev, N. I.; Faraon, A.; Arbabi, A.; Yu, N.; Chanda, D.; Crozier, K. B.; Kildishev, A. V.; Wang, H.; Yang, J. K. W.; Valentine, J. G.; Genevet, P.; Fan, J. A.; Miller, O. D.; Majumdar, A.; Fröch, J. E.; Brady, D.; Heide, F.; Veeraraghavan, A.; Engheta, N.; Alù, A.; Polman, A.; Atwater, H. A.; Thureja, P.; Paniagua-Dominguez, R.; Ha, S. T.; Barreda, A. I.; Schuller, J. A.; Staude, I.; Grinblat, G.; Kivshar, Y.; Peana, S.; Yelin, S. F.; Senichev, A.; Shalaev, V. M.; Saha, S.; Boltasseva, A.; Rho, J.; Oh, D. K.; Kim, J.; Park, J.; Devlin, R.; Pala, R. A. Roadmap for Optical Metasurfaces. *ACS Photonics* **2024**, *11* (3), 816–865. https://doi.org/10.1021/acsphotonics.3c00457.

(17) Liu, W.; Li, Z.; Cheng, H.; Chen, S. Dielectric Resonance-Based Optical Metasurfaces: From Fundamentals to Applications. *iScience* **2020**, *23* (12), 101868. https://doi.org/10.1016/j.isci.2020.101868.

(18) Makarenko, M. O.; Burguete-Lopez, A.; Rodionov, S.; Wang, Q.; Getman, F.; Fratalocchi, A. Experimental Machine Learning for Aperiodic Wafer-Scale Photonics Inverse Design. In *Machine Learning in Photonics*; Ferranti, F., Hedayati, M. K., Fratalocchi, A., Eds.; SPIE, 2024; p 25. https://doi.org/10.1117/12.3017331.

(19) Neuman, T.; Alonso-González, P.; Garcia-Etxarri, A.; Schnell, M.; Hillenbrand, R.; Aizpurua, J. Mapping the near Fields of Plasmonic Nanoantennas by Scattering-Type Scanning near-Field Optical Microscopy. *Laser & Photonics Reviews* **2015**, *9* (6), 637–649. https://doi.org/10.1002/lpor.201500031.

(20) Bouchal, P.; Dvořák, P.; Babocký, J.; Bouchal, Z.; Ligmajer, F.; Hrtoň, M.; Křápek, V.; Faßbender, A.; Linden, S.; Chmelík, R.; Šikola, T. High-Resolution Quantitative Phase Imaging of Plasmonic Metasurfaces with Sensitivity down to a Single Nanoantenna. *Nano Letters* **2019**, *19* (2), 1242–1250. https://doi.org/10.1021/acs.nanolett.8b04776.

(21) Khadir, S.; Andrén, D.; Verre, R.; Song, Q.; Monneret, S.; Genevet, P.; Käll, M.; Baffou, G. Metasurface Optical Characterization Using Quadriwave Lateral Shearing Interferometry. *ACS Photonics* **2021**, *8* (2), 603–613. https://doi.org/10.1021/acsphotonics.0c01707.

(22) Babocký, J.; Křížová, A.; Štrbková, L.; Kejík, L.; Ligmajer, F.; Hrtoň, M.; Dvořák, P.; Týč, M.; Čolláková, J.; Křápek, V.; Kalousek, R.; Chmelík, R.; Šikola, T. Quantitative 3D Phase Imaging of Plasmonic Metasurfaces. *ACS Photonics* **2017**, *4* (6), 1389–1397. https://doi.org/10.1021/acsphotonics.7b00022.

(23) Faßbender, A.; Babocký, J.; Dvořák, P.; Křápek, V.; Linden, S. Direct Phase Mapping of Broadband Laguerre-Gaussian Metasurfaces. *APL Photonics* **2018**, *3* (11), 110803. https://doi.org/10.1063/1.5049368.

(24) Badloe, T.; Lee, J.; Seong, J.; Rho, J. Tunable Metasurfaces: The Path to Fully Active Nanophotonics. *Advanced Photonics Research* **2021**, *2* (9). https://doi.org/10.1002/adpr.202000205.

(25) Cui, T.; Bai, B.; Sun, H. Tunable Metasurfaces Based on Active Materials. *Advanced Functional Materials* **2019**, *29* (10). https://doi.org/10.1002/adfm.201806692.

(26) Kepič, P.; Ligmajer, F.; Hrtoň, M.; Ren, H.; Menezes, L. de S.; Maier, S. A.; Šikola, T. Optically Tunable Mie Resonance $VO_2$ Nanoantennas for Metasurfaces in the Visible. *ACS Photonics* **2021**, *8* (4), 1048–1057. https://doi.org/10.1021/acsphotonics.1c00222.

(27) Chmelik, R.; Slaba, M.; Kollarova, V.; Slaby, T.; Lostak, M.; Collakova, J.; Dostal, Z. The Role of Coherence in Image Formation in Holographic Microscopy; 2014; pp 267–335. https://doi.org/10.1016/B978-0-444-63379-8.00005-2.

(28) Jiang, Q.; Liu, J.; Li, J.; Jing, X.; Li, X.; Huang, L.; Wang, Y. Multiwavelength Achromatic Metalens in Visible by Inverse Design. *Adv Opt Mater* **2023**, *11* (15), 2300077. https://doi.org/10.1002/ADOM.202300077.

(29) Yin, Y.; Jiang, Q.; Wang, H.; Liu, J.; Xie, Y.; Wang, Q.; Wang, Y.; Huang, L. Multi-Dimensional Multiplexed Metasurface Holography by Inverse Design. *Advanced Materials* **2024**, *36* (21), 2312303. https://doi.org/10.1002/ADMA.202312303.

(30) Ji, A.; Song, J.-H.; Li, Q.; Xu, F.; Tsai, C.-T.; Tiberio, R. C.; Cui, B.; Lalanne, P.; Kik, P. G.; Miller, D. A. B.; Brongersma, M. L. Quantitative Phase Contrast Imaging with a Nonlocal Angle-Selective Metasurface. *Nature Communications* **2022**, *13* (1), 7848. https://doi.org/10.1038/s41467-022-34197-6.

(31) Song, Q.; Khadir, S.; Vézian, S.; Damilano, B.; de Mierry, P.; Chenot, S.; Brandli, V.; Laberdesque, R.; Wattellier, B.; Genevet, P. Printing Polarization and Phase at the Optical

Diffraction Limit: Near- and Far-Field Optical Encryption. *Nanophotonics* **2020**, *10* (1), 697–704. https://doi.org/10.1515/nanoph-2020-0352.

(32) Chen, Y.-C.; Zeng, Q.-C.; Yu, C.-Y.; Wang, C.-M. General Case of the Overall Phase Modulation through a Dielectric PB-Phase Metasurface. *OSA Continuum* **2021**, *4* (12), 3204. https://doi.org/10.1364/OSAC.441987.

(33) Fan, C.-Y.; Lin, C.-P.; Su, G.-D. J. Ultrawide-Angle and High-Efficiency Metalens in Hexagonal Arrangement. *Scientific Reports* **2020**, *10* (1), 15677. https://doi.org/10.1038/s41598-020-72668-2.

(34) Spägele, C.; Tamagnone, M.; Kazakov, D.; Ossiander, M.; Piccardo, M.; Capasso, F. Multifunctional Wide-Angle Optics and Lasing Based on Supercell Metasurfaces. *Nature Communications* **2021**, *12* (1), 3787. https://doi.org/10.1038/s41467-021-24071-2.